\documentclass[    ,final            
 ]
  {aipproc}

\layoutstyle{8x11double}

\begin{document}

\title{The Boson peak and the phonons in glasses}

\author{S.~Ciliberti}{
  address={INFM UdR
Roma1, Universit\'a di Roma ``La Sapienza'', and Center for
Statistical Mechanics and Complexity (SMC), P.le A. Moro 2, I-00185
Roma, Italy}
}

\author{T.~S.~Grigera}{
  address={Centro di Studi e
Ricerche ``Enrico Fermi'', via Panisperna 89/A, I-00184 Roma, Italy}
}

\author{V.~Mart\'{\i}n-Mayor{}}{
  address={Departamento de Fisica Teorica I, Universidad Complutense
de Madrid, Madrid 28040, Spain; Instituto de Biocomputaci\'on y
F\'{\i}sica de Sistemas Complejos (BIFI). Universidad de Zaragoza,
50009 Zaragoza, Spain.}
}

\author{G.~Parisi}{
  address={INFM UdR
Roma1, Universit\'a di Roma ``La Sapienza'', and Center for
Statistical Mechanics and Complexity (SMC), P.le A. Moro 2, I-00185
Roma, Italy}
}

\author{P.~Verrocchio}{
  address={Departamento de Fisica Teorica I, Universidad Complutense
de Madrid, Madrid 28040, Spain; Instituto de Biocomputaci\'on y
F\'{\i}sica de Sistemas Complejos (BIFI). Universidad de Zaragoza,
50009 Zaragoza, Spain.}
}

\begin{abstract}
Despite the presence of topological disorder, phonons seem to exist
also in glasses at very high frequencies (THz) and they remarkably
persist into the supercooled liquid.  A universal feature of such a
systems is the Boson peak, an excess of states over the standard Debye
contribution at the vibrational density of states. Exploiting the
euclidean random matrix theory of vibrations in amorphous systems. we
show that this peak is the signature of a phase transition in the
space of the stationary points of the energy, from a minima-dominated
phase (with phonons) at low energy to a saddle-point dominated phase
(without phonons). The theoretical predictions are checked by means of
numeric simulations.
\end{abstract}

\maketitle

\section{Introduction}

$X$-ray and neutron scattering techniques allow to obtain very
detailed physical insight into the high-frequency (0.1--10$\,$THz)
vibrational dynamics of supercooled liquids and glasses. Within this
range of frequencies their spectra reveal several universal
properties~\cite{exp}, related with the presence of sound-like
excitations even for momenta $p$ of the same order of magnitude of
$p_0$, the first maximum of the static structure factor (typically
corresponding to wave numbers of a few $nm^{-1}$).  This
\emph{high-frequency sound} is revealed as Brillouin-like peaks in the
Thz region of the dynamic structure factor. An accessible quantity to
experiments is the vibrational density of states (VDOS), $g(\omega)$,
whose most striking feature is the presence of an excess of states
over the Debye $\omega^2$ law in the ``low'' frequency region,
(i.e.where the dispersion relation is linear, but still in the Thz
region)\cite{debate}. This excess of states is seen as a peak when
plotting $g(\omega)/\omega^2$ and has been named {\em Boson peak} (BP)
\footnote{There exist alternative ways of defining the boson peak from
experiments, for example as a peak in Raman scattering data or as a
peak in the difference between the observed VDOS of the glass and that
of the corresponding crystal}. The peak position $\omega_{BP}$ usually
shifts to lower frequency on heating~\cite{bpvsheat}, except for the
case of silica~\cite{bpsilica1}. In this material the shift is seen on
lowering the density~\cite{bpsilica2}.

Due to its universality, the relevant physics underlying the Boson
peak can be hopefully captured by some simple model.  Furthermore,
several recent numerical simulations have shown that a model of
harmonic vibrations is wholly adequate to describe this frequency
range~\cite{harmonic} and that anharmonicity need not be
invoked. Given the presence of well formed local structures (SiO$_2$
tetrahedra, for instance) a natural approximation is to consider that
the oscillation centers form a crystalline structure, the disorder in
the atomic positions being mimicked by randomness in their interaction
potential~\cite{schirmacher98,taraskin} (disordered lattice
models~\cite{elliott}).  The main drawback of such a models is that
they dramatically underestimate the scattering of sound
waves~\cite{mayor00}. A different approach studies vibrations around a
topologically disordered~\cite{elliott} (liquid like) structure. It is
followed by two different theories: modified mode-coupling
theory~\cite{mct} (which is not limited to harmonic excitations), and
euclidean random matrix theory (ERMT)~\cite{ermZee,erm2,erm2bis}. ERMT
owes its name to the fact that it formulates the vibrational problem
as random matrix problem~\cite{metha}. The matrices involved are
called Euclidean random matrices~\cite{erm}, and their study has
required the development of new analytical tools.  Both MCT and ERMT
predict an enhanced scattering of sound waves as compared to
disordered crystals.

On the other hand, even within the harmonic framework the nature of
the extra low frequency modes giving rise to the BP is still an open
point. At a qualitative level, the frequency $\omega_{BP}$ is close to
the Ioffe-Regel \cite{iofferegel} frequency $\omega_{IR}$, suggesting
the possibility that the excess BP modes are
localized~\cite{alexander89}. However, numerical simulations have
shown that the localization edge is at frequencies greater than
$\omega_{BP}$ and $\omega_{IR}$~\cite{diffusoni}. The Ioffe-Regel
criterion signals rather a crossover to a region where the harmonic
excitations are not longer propagating, due to the strong interaction
with the disorder. We call these modes {\em glassons} (since they do
not propagate but ``diffuse'', they have also been called
\emph{diffusons}~\cite{diffusoni}). A large bump of glassons is
generally found around the Ioffe-Regel frequency, due to the
flattening of the dispersion relation. This can be considered as the
glass counterpart of the van Hove singularity of
crystals~\cite{taraskin,erm3}. All the recently proposed theoretical
frameworks predict that this peak of glassons should move to lower
frequencies when approaching an instability transition, where negative
eigenvalues (imaginary frequencies) appear.  The aim of the paper is
showing that the ERM theory makes sharp predictions about the values
of universal critical exponents describing the approach to this
singularity and comparing them with numeric results. The emerging
scenario describes the BP modes as given by the hybridization between
the phonons and the low-energy tail of the glasson peak which softens
when the system approaches the instability
transition~\cite{nature,vetto}.

\section{The Euclidean Random Matrix Theory}

The starting approximation is that particles can only oscillate around
fixed random positions, so that the position of particle $i$ at time
$t$ is ${\bf x}_i(t) = {\bf x}^{eq}_i + {\bf \varphi}_i(t)$; the ${\bf
x}^{eq}_i$ are quenched equilibrium positions (whose distribution must
be specified) and ${\bf \varphi}_i(t)$ are the displacements. Hence
the Hamiltonian is
\begin{equation} 
H\left[{\bf x}\right]= \sum_{i,j}^{1,N} V({\bf x}_i-{\bf x}_j) \simeq
\frac 12\sum_{i,j}^{1,N}\sum_{\mu,\nu}^{1,3} M_{i\mu,j\nu}[{\bf
x}^{eq}]\varphi_i^\mu\varphi_j^\nu
\end{equation} 
where the dynamical matrix M is an Euclidean Random Matrix:
\begin{equation} 
M_{i\mu,j\nu}[{\bf x}^{eq}]\equiv
- f_{\mu\nu}({\bf x}_i^{eq}-{\bf x}_j^{eq})
\ +\ 
\delta_{ij}\sum_{k=1}^N f_{\mu\nu}({\bf x}_i^{eq}- {\bf x}_k^{eq}),
\end{equation} 
with $f_{\mu\nu}({\bf x})\equiv\partial_{\mu\nu} V({\bf
x})$. 

In the one-excitation approximation the dynamic structure factor is
\begin{equation} 
S^{(1)}({\bf p},\omega)= \frac{k_{\mathrm B} T}{m\omega^2} \overline{
\sum_n \left| \sum_i {\bf p}\cdot{\bf e}_{n,i} e^{ i{\bf p} \cdot {\bf
x}^{eq}_i } \right|^2 \delta(\omega -\omega_n)},
\end{equation} 
where ${\bf e}_n$ are the eigenvectors of the dynamical matrix and
$\omega_n$ its eigenfrequencies (= square root of eigenvalues).  The
overline means average over the disordered quenched positions, whose
distribution $P[ {\bf x}^{eq} ]$ has to be specified.  The density of
states (VDOS) is obtained in the limit of large momenta:
\begin{equation}
g(\omega)=\lim_{p\to\infty} {m\omega^2 \over k_B T p^2} 
S^{(1)}(p,\omega).   
\label{s-dos}
\end{equation}
We can obtain $S^{(1)}({\bf p},\omega)$ through the resolvent 
$G({\bf p},z)$:
\begin{eqnarray}
G_{\mu\nu}({\bf p},z) \equiv \frac{1}{N} \sum_{jk}
\overline{ 
e^{i {\bf p} \cdot (
{\bf x }^{\mathrm eq}_j-{\bf x}^{eq}_k ) }
\left[ {1 \over z-M} \right]}_{j\mu,k\nu} \nonumber \\
\equiv G_L(p,z)\frac{ p_\mu p_\nu}{p^2} + 
G_T(p,z)\left(\delta_{\mu\nu}-\frac{ p_\mu p_\nu}{p^2}\right)   \label{s-res}
\end{eqnarray}
separating the axial tensor in a longitudinal term and a transversal
one.  The dynamic structure factor is then:
\begin{equation}
S^{(1)}({\bf p},\omega)= 
- \frac{2 k_\mathrm{B} T p^2}{\omega\pi} 
\mathrm{Im}\, G_{L}({\bf p},\omega^2 +{\mathrm i} 0^+).
\label{sdiq}
\end{equation}
A transverse dynamic structure factor (not measurable in experiments)
can be defined in an analogous way. However, a most important and
general result is that for $p\to\infty$ the resolvent becomes
isotropic:
\begin{equation}
G^\infty_{\mu\nu}(z) = \frac{1}{N} \sum_{j}
\overline{
\left[ {1 \over z-M} \right]}_{j\mu,j\nu} 
= \delta_{\mu\nu} \frac 1N \overline{ \textrm{Tr}\: [z-M]^{-1} }.
\end{equation}
So both longitudinal and transverse structure factors tend to a common
limit (the VDOS, see eq.~\ref{s-dos}) at infinite momentum.  
\footnote{Conseguently, both the dispersion relations saturate at the
same value. However due to the broadening of the line, they are rather
ill-defined when $\omega \sim \omega_{IR}$} Leaving the potential
$V(r)$ unspecified and taking the simplified case $P[{\bf
x}^{eq}]=1/V^N$ ($V$ being the volume), one finds that:
\begin{equation}
 G_{\mu \nu}({\bf p},z)=\left[\frac{1}{z-\rho \hat f(0)+ \rho \hat
 f({\bf p}) - \Sigma ({\bf p},z)}\right]_{\mu \nu},
\end{equation}
The self-energy $\Sigma ({\bf p},z)$ is a matrix with the
standard form
\begin{equation}
\Sigma_{\mu\nu}({\bf p},z)=\Sigma_L(p) \frac{p_\mu p_\nu}{p^2} +
\Sigma_T(p) \left(\delta_{\mu \nu} - \frac{p_\mu p_\nu}{p^2}
\right).
\end{equation}
which vanishes at $\rho=\infty$ and that can be computed in a series
expansion in $1/\rho$. The main point is that the sum of all the
infinite diagrams obtained recursively starting from this
next-to-leading order result gives a self-consistent integral
equation\cite{vetto}:
\begin{equation} 
\Sigma_{\mu\nu}({\bf p},z)=\frac{1}{{\rho}} \int\frac{d^3 q}{(2\pi)^3} 
V_{\mu\lambda}({\bf q},{\bf p})G_{\lambda\sigma}({\bf q},z)
V_{\sigma\nu}({\bf q},{\bf p}).
\label{eqint}
\end{equation}
where the vertices have the form $V_{\mu\nu}({\bf q},{\bf p})=\rho
(\hat f_{\mu\nu}({\bf q}) - \hat f_{\mu\nu}({\bf p}-{\bf q}))$.  Let
us remark that the self energy renormalizes the dispersion relations
and gives a finite width to the Brillouin peaks:
\begin{eqnarray}
\omega^2_{L,T}(p)&=&
\left(\omega^0_{L,T}\right)^2 (p)+ \textrm{Re} \:
\Sigma_{L,T}(p,\omega_{L,T}(p)),
\nonumber \\
\Gamma_{L,T}(p)&=&\textrm{Im} \:
\Sigma_{L,T}(p,\omega_{L,T}(p))/\omega_{L,T}(p).
\label{allargamento}
\end{eqnarray}

The correlations between the equilibrium positions of the particles
can be taken into account quite easily at the level of the {\em
superposition approximation} in the above approach. The results
derived above for the case without correlations are translated to the
correlated case by replacing the functions $f({\bf x})$ by
$g^{(2)}({\bf x}) f({\bf x})$. In this way the usual power law
divergence of the pair potential for $|{\bf x}|\to 0$ is balanced by
the exponential behaviour of the pair distribution function, and this
ensures the existence of the Fourier transform of the product $f({\bf
x})g^{(2)}({\bf x})$.

\subsection{The phase transition}

From equation~(\ref{eqint}) it is possible to derive a few analytic
model-independent results about the arising of the Boson Peak. These
results are expressed in form of scaling laws, whose exponents are
predicted in this approximation. Simulations (see below) and
experiments will allow to clarify the dependence of the exponents on
the approximation. The VDOS can be obtained from
\begin{equation}
g(\omega)=-\frac{2\omega}\pi\ \textrm{Im}\ G^\infty(\omega^2+i0^+) ,
\label{dos}
\end{equation}
where $z=\omega^2+i0^+$ and $G^\infty(z)\equiv\lim_{p\to\infty} G({\bf p},z)$.

Hence one have to solve the integral equation~(\ref{eqint}) in the
$p\to\infty$ limit:
\begin{equation}
\frac{1}{G^\infty(z)} = z - \rho \hat f(0)-
\rho A G^\infty(z)
- \rho\int\!\!\frac{d^3q}{(2\pi)^3}\, \hat f^2({\bf q})G({\bf q},z) 
\label{asint}
\end{equation}
where $G^\infty(z)$, $A \equiv (2\pi)^{-3} \int d^3q \hat f^2({\bf
q})$ and the last term are matrices proportional to the identity.

The solution of the above integral equation yields a VDOS which
contains both the phonons, since $g(\omega) \propto \omega^2$ at
$\omega \to 0$, and the extended but not propagating glassons,
described by a semicircle with center at $\omega= \rho \hat f(0)$ and
radius $2\sqrt{\rho A}$~\cite{erm3}. If we limit (for pedagogical
purposes) to the case where the VDOS changes because of changes in the
density, the key point is the existence of a phase transition in the
space of the eigenvalues of the Hessian matrix.  In fact $G^\infty(0)$
develops an imaginary part when $\rho<\rho_c$ ($\rho_c$ being a
critical density), then the transition separates the stable phase (all
positive eigenvalues) and the unstable phase (negative and positive
eigenvalues). The glassons are the modes which move towards the
negative zone of the spectrum (hybridizing the phonons) when
approaching the transition. The order parameter is $\varphi =
-\textrm{Im}\,G^\infty(i0^+)$ which vanishes as $\varphi\sim
|\Delta|^\beta$, with $\beta=1/2$ and $\Delta \equiv \rho-\rho_c$

The relation with the Boson Peak becomes quite clear when one writes
down the VDOS in the stable phase arising from the theory without any
reference to the control parameter, which then does not need to be
$\rho$. In fact, one has
\begin{equation}
g(\omega,\Delta)=\omega^\gamma h(\omega\Delta^{-\rho}), 
\quad h(x)\sim \left\{
\begin{array}{lcr}
x^{2-\gamma}&&x\ll 1\\
\textrm{const.}&&x\gg 1
\end{array}
\right. ,
\end{equation}
with $\Delta$ defined in terms of an arbitrary control parameter.The
ERMT (in the cactus approximation) predicts $\rho = 1$, $\gamma =
3/2$. Hence it exist a crossover frequency (in the region where the
dispertion relation is still linear) between a $\omega^2$ and a
$\omega^{\gamma}$ region. We shall identify that with the BP frequency
$\omega_{BP}$. This implies that $\omega_{BP}\sim \Delta^\rho$ and
$g(\omega_{BP},\Delta)/\omega_{BP}^2\sim\Delta^{-\eta}$, with
\begin{equation}
\eta=\rho(2-\gamma)\; .
\end{equation}
Let us note that the BP is indicated from a peak in the function
$g(\omega)/\omega^2$m not in $g(\omega)$. Summarizing, according to
ERMT the BP frequency moves linearly toward $0$ when approaching the
transition (from the stable side) and its height diverges as a power
law whose exponent is $\eta=1/2$. Eq.~(\ref{s-dos}) shows that at the
level of the one-phonon approximation it can also be detected in the
large $p$ limit of the dynamic structure factor $S(p,\omega)$. 

\section{Boson Peak in a gaussian model}

\begin{figure}
  \includegraphics[height=.3\textheight]{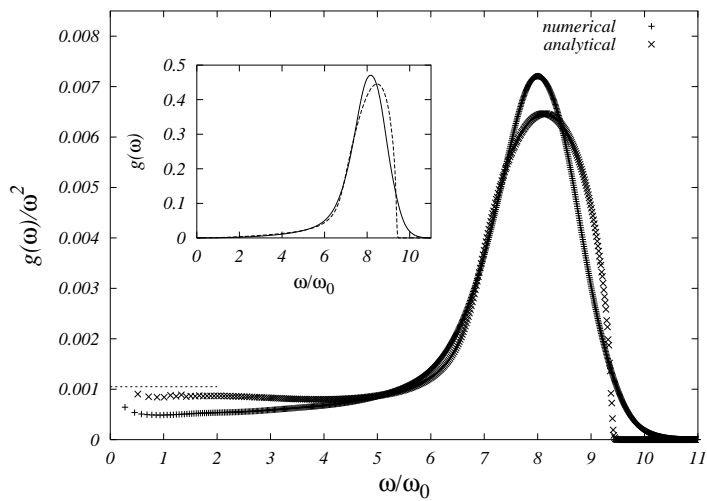} 
  \caption{The VDOS g($\omega$) as a function of eigenfrequencies
  divided by the Debye behaviour $\omega^2$ for $\rho=4>\rho_c$, both
  numerical (obtained via the method of moments) and analytical. In
  the inset, we show g($\omega$) vs. $\omega$.}  
  \label{confronto}
\end{figure}

In order to confirm that the saddle-phonon transition described by the
Euclidean Random Matrix theory is not an artifact of the approximation
involved (cactus resummation), we solved numerically the cactus
equation for the case where $f(p)$ has a Gaussian form and compare
with direct numerical results for the same model\cite{vetto}. The model is
described by
\begin{eqnarray}
\hat f_{\mu\nu}({\bf p}) &=& \hat f_L(p)\frac{p_\mu p_\nu}{p^2} +
\hat f_T(p) \left(\delta_{\mu\nu}-\frac{p_\mu p_\nu}{p^2}\right), 
\nonumber \\
\hat f_{L,T}(p)&=&\left(\frac{2\pi}{\sigma_0^2}\right)^{3/2}
\exp{(-p^2/2\sigma^2_{L,T})} .
\label{model}
\end{eqnarray}
This choice for $\hat f(p)$ is mainly due to its simplicity. However
the behaviour of the Boson peak close to the saddle-phonon transition
have to be independent of the details of the model.  Moreover the
superposition approximation takes $\hat f_{\mu\nu}(p)={\cal F}[g(r)
v_{\mu\nu}(r)]$ yielding a $\hat f_{L,T}(p)$ finite at $p=0$, like in
the Gaussian model.  We shall consider various values of the density,
which is here the control parameter, comparing the analytical (cactus)
results with the numerical spectra and dynamic structure factor
obtained from the method of moments~\cite{mom}. In the high density
regime the approximations used in deriving the integral
equation~(\ref{asint}) are quite good since the analytic solution
reproduces the numerical spectrum (and in particular the Debye
behaviour) rather accurately (fig.\ref{confronto}).

\begin{figure}
  \includegraphics[height=.3\textheight]{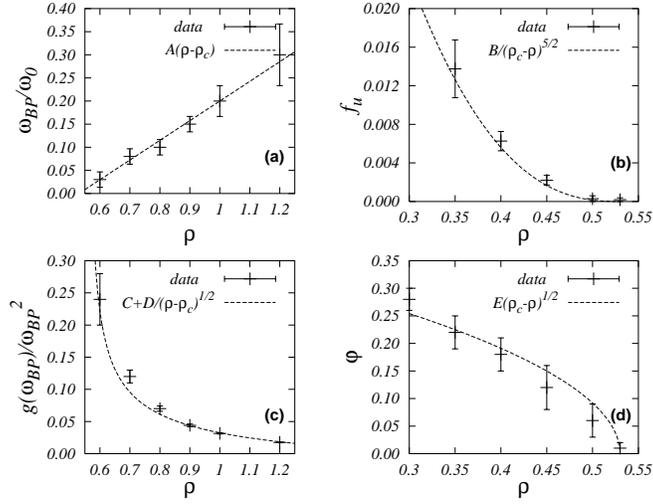}
   \caption{Numeric results. The critical density in the following
   fits has been fixed to $\rho_c=0.54$ and capital letters are the
   fitting parameters. {\bf(a)}: The position of Boson peak as a function
   of the density near the critical point. $\omega_{BP}$ vanishes
   linearly in $\Delta=\rho-\rho_c$. {\bf (b)}: The fraction of unstable
   modes vanishes as $(\rho_c-\rho)^{2.5}$, thus yielding
   $\gamma=3/2$. {\bf (c)}: The height of the BP, defined by
   $g(\omega_{BP})/\omega_{BP}^2$, diverges as $\Delta^{-\eta}$, with
   $\eta \sim 1/2$. {\bf (d)}: The order parameter $\varphi\equiv -\mbox{Im}\,
   G^{\infty}(0)$ vanishes as $(\rho_c-\rho)^\beta$, with $\beta \sim
   1/2$.}  
   \label{gauss}

\end{figure}

However, the crucial check regards the exponents of the transition.
Figs.~\ref{gauss}a and \ref{gauss}c show that the position of the BP
is linear with respect to $\Delta\equiv (\rho-\rho_c)$ and that the
height of BP diverges as $\Delta^{-1/2}$. This confirms the
theoretical predictions $\nu=1$ and $\eta=1/2$. In Fig.~\ref{gauss}b
we determine the value of $\gamma$ by studying the fraction of
unstable modes. In fact, in the region of parameters where $\rho <
\rho_c$ the fraction of unstable modes, defined as
$f_u=\int_{-\infty}^0 g_\lambda(\lambda)d\lambda$, is given by
\begin{equation}
f_u(\Delta)=\int_0^\infty\!\! d\omega\, \omega^\gamma \tilde
g(\omega/|\Delta|) \sim|\Delta|^{1+\gamma} .
\end{equation}
We find numerically (Figs.~\ref{bp_num}b) that
$f_u\sim(\rho_c-\rho)^{5/2}$, i.e.\ $\gamma=3/2$. Finally, the order
parameter $\varphi$ vanishes as $(\rho_c-\rho)^\beta$ with $\beta=1/2$
(Fig.~\ref{gauss}d).

Hence our analytic treatment based on euclidean random matrix theory
describe quite well the vibrational features of simple topologically
disordered systems\cite{vetto}. The following step is understanding to
what degree of accurateness ERMT could describe the high frequency
properties of more realistic systems\cite{nature}.

\section{Boson Peak in a fragile glass}

Starting from the hypothesis that the Thz region of supercooled
liquids and glasses can be described in terms of purely harmonic
excitations, the origin of the Boson peak in glasses can be understood
if we consider the ensemble of generalized inherent structures (GIS).
For each equilibrium configuration the associated GIS is the nearest
stationary point of the Hamiltonian.  If we start from an equilibrium
configuration at low temperature, the GIS is a local minimum, and it
coincides with the more frequently used inherent structures (IS)
\cite{INM}(i.e.\ the nearest minimum of the Hamiltonian). On the
contrary, if we start from high temperature, the GISs are saddle
points. In the GIS ensemble there is a sharp phase transition
separating these two regimes.  It takes place in glass-forming
liquids~\cite{saddles} at the Mode Coupling temperature~\cite{tmc}
($T_{MC}$), above which liquid diffusion is no longer ruled by rare
``activated'' jumps between ISs but by the motion along the unstable
directions of saddles.  Phonons are present in the spectrum of the
VDOS in the low temperature phase (IS dominated) but are absent in the
saddle phase.  The key point is that the minima obtained starting from
configurations below $T_{MC}$ and the saddles obtained starting above
$T_{MC}$ join smoothly at $T_{MC}$. Thus we can study GIS as a single
ensemble parametrized by their energy \cite{saddles}.

Since this transition separates a phase where all the eigenvalues are
positive from another one where even negative eigenvalues exist, we
expect that ERMT is able to describe correctly this phenomenum. Hence
we measured numerically the values of some exponents predicted by the
theory in a simple model of a fragile glass~\cite{nature}. We
simulated a soft-spheres binary mixture~\cite{Bernu87} in the stable
(phonon) phase with the Swap Monte Carlo algorithm
\cite{GrigeraParisi01}, and computed the VDOS of the ISs obtained
starting from equilibrium configurations at temperatures below
$T_{MC}$\footnote{At very low $T$, where equilibrium is not achieved,
runs were followed until $e_{IS}$ got very close to its asymptotic
value}.

\begin{figure}
  \includegraphics[height=.3\textheight,angle=270]{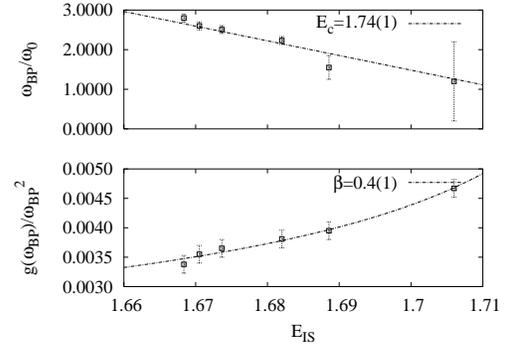}
  \caption{Scaling of the position, $\omega_{BP}$, and height of the
  Boson peak near the saddle-phonon transition (energies and
  frequencies in units of $\epsilon$ and $\omega_0$
  respectively). {\bf Top} $\omega_{BP}$ is linear in the control
  parameter $e_{IS}$ and vanishes at $e_{IS}=e_c=1.74(1) \,
  \epsilon$. {\bf Bottom} The height of the Boson peak diverges as a
  power law with exponent $\beta \sim 0.4$. Height and position of the
  BP were obtained by fitting a parabola to the peak of
  $g(\omega)/\omega^2$. {\it Reprinted with permission from
  Nature~\cite{nature}, Copyright (2003) Macmillan Magazines Limited}}
  \label{bp_num}
\end{figure}

In Fig.\ref{bp_num} we show that the theoretical predictions agree
with the numerical data. Taking the IS's energy as the relevant
parameter for describing the spectral properties, one has $\Delta =
e_c-e_{IS}$, $e_{IS}$ being the energy of the ISs and $e_c$ the
critical value. In fact, plotting $g(\omega)/\omega^2$ a peak is
clearly identified, which is seen to grow in height and shift to lower
frequency on rising the IS's energy. Using all the spectra for which
the peak position can be clearly identified, we find that the
relationship between $\omega_{BP}$ and the energy of the IS is linear
(Fig. 3a).  The energy at which $\omega_{BP}$ becomes zero, $e_c$, is
found from a linear fit as $e_c=1.74 \, \epsilon$ ($\epsilon$ is the
energy scale), quite close to the value where the GIS stop to be
minima (IS) and become saddles \cite{saddles}. As for the height of
the peak (Fig.~3b), the results are compatible with a power-law
divergence. Fixing $e_c$ at the value $1.74 \, \epsilon$ arising from
the linear fit of $\omega_{BP}$ {\sl vs.}  $e_{IS}$, a power-law fit
yields an exponent $\beta=0.40(15)$, while if one fixes the exponent
at $\beta=1/2$, then the critical value is $e_c=1.752(2) \,
\epsilon$. Thus the numerical data are compatible with the
theoretically predicted scaling, although we have not been able to
work close to the critical point, and thus cannot get a great accuracy
on the critical exponents or the critical point.

\section{Conclusions}

In summary, we have shown that the saddle (negative
eigenvalues)-phonon (no negative eigenvalues) transition, a common
feauture of vibrating topologically disordered systems, is well
described by the euclidean random matrix theory. It provides a
coherent scenario for the arising of a Boson Peak, which results from
the hybridization of acoustic modes with high-energy modes that soften
upon approaching the transition.  Hence we applied the theory to
describe the saddle-phonon transition and the BP in supecooled
liquids, comparing the predicted scaling laws with the numeric results
obtained for a simple fragile glass former. The agreement found is
quite encouraging. The present discussion applies to experiments as
long as one is in the regime where the inverse frequency is much
larger than the structural relaxation time, when the harmonic
approximation makes sense.  We expect that the saddle-phonon
transition point of view will be able to bridge the realms of
experiment and numerical studies of the energy landscape.  As a matter
of fact, a recent experiment on the poly(methyl methacrylate) (PMMA)
glass gave the first experimental confirmation of the ERMT
predictions\cite{Duval03}.

\begin{theacknowledgments}

V.M.-M. is a {\em Ram\'on y Cajal} research fellow (MCyT,
Spain). P.V. was supported through the European Community's Human
Potential Programme under contract HPRN-CT-2002-00307, DYGLAGEMEM.

\end{theacknowledgments}


\bibliographystyle{aipproc}   


\end{document}